\documentclass[sigconf]{acmart}
\renewcommand\footnotetextcopyrightpermission[1]{} 
\usepackage{booktabs} 

\begin{document}
\title{Sloth Search System at the Video Browser Showdown 2018 -- Final Notes}

\author{Nattachai Watcharapinchai}
\orcid{0000-0001-6691-1243}
\affiliation{%
  \institution{National Electronics and Computer Technology Center (NECTEC)}
  \country{Thailand}}
\email{nattachai.wat@nectec.or.th}
\author{Sitapa Rujikietgumjorn}
\orcid{0000-0002-5826-3319}
\affiliation{%
  \institution{National Electronics and Computer Technology Center (NECTEC)}
  \country{Thailand}
}
\email{sitapa.ruj@nectec.or.th}

\author{Sanparith Marukatat}
\affiliation{%
  \institution{National Electronics and Computer Technology Center (NECTEC)}
  \country{Thailand}
}
\email{sanparith.mar@nectec.or.th}

\begin{abstract}
This short paper provides further details of the Sloth Search System, which was developed by the NECTEC team for the Video Browser Showdown (VBS) 2018.  
\end{abstract}

\keywords{Image/Video Retrieval System, Video Browsing Showdown 2018}

\maketitle

\section{System Overview}

The NECTEC team is a first-time participant in the VBS competition in 2018. We have presented a video retrieval system as an interactive browsing tool with a simple interface. The main system is designed to support two types of retrieval modes which are text-based retrieval and sketch-based retrieval~\cite{Sloth2017}. All features can be summarized as follows:

\begin{itemize}
\item indexing with high level concepts of 16,429 concepts from three convolutional neural network (CNN) models, 
\item indexing with eight dominant color masks and the top 10 object localizations using masks of their bounding boxes,
\item using hashing technique to search for color masks and object localization masks,
\item reducing the retrieval time by using the Elasticsearch system for a full text search, 
\item reducing the retrieval time by using the Radis system for sketch-based search, 
\item viewing all keyframes in a related video shot,
\item reranking the results by using grouping with video shot.
\item combining the text-based and sketch-based search by using weighted technique.
\item two display modes for the retrieval results: sort all frame by the similarity score and group the frames by the video id. 
\end{itemize}


\section{Final System Changes}
The system described in our VBS2018 paper~\cite{Sloth2017} has been further improved after paper submission. The details of these changes are described in this section.

\begin{figure}[h]
\begin{center}
\includegraphics[width=0.8\linewidth]{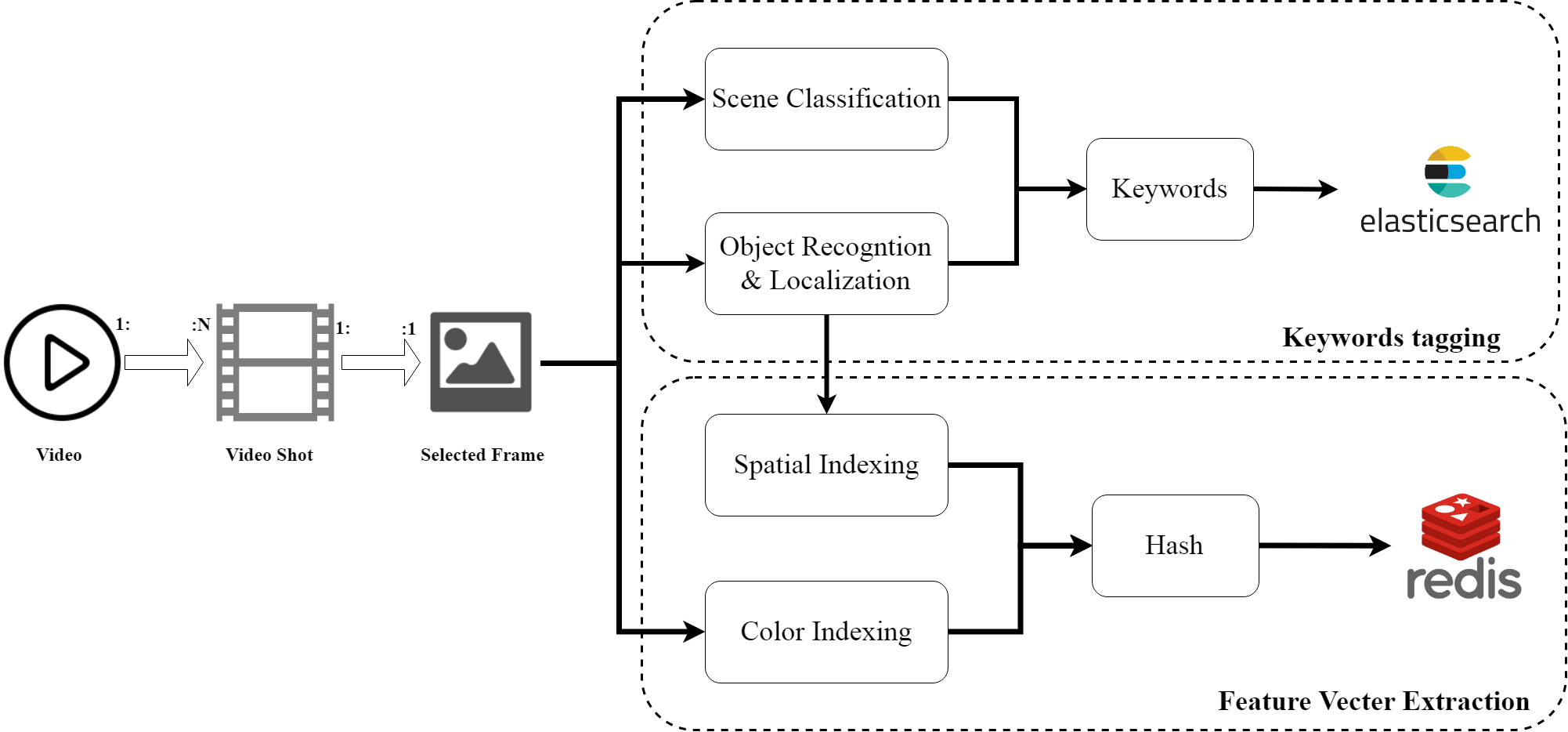}
\end{center}
\caption{Overview of our indexing process.}
\label{fig:overview}
\end{figure}

The final system was developed using Elasticsearch~\cite{Elasticseach2015} together with Radis~\cite{Redis2013} as shown in Fig.~\ref{fig:overview} to improve the retrieval time. 
For indexing process, instead of using every frames in the whole dataset, we used 335,944 keyframes provided with the TRECVID dataset~\cite{2017trecvidawad}. 
For each keyframe, the concept labels, color feature, and object spatial vectors were extracted. 
We used the CNNs to extract objects, scenes, and image captions as concept labels. 
The objects were recognized and localized using the Faster R-CNN, which was trained on the Open Images dataset~\cite{openimages} with 545 object classes. 
Scenes were extracted using Alexnet trained on the Place365 dataset\cite{zhou2017places}. 
And we used image caption model or im2txt~\cite{Vinyals2017} to generate image captions. These extracted labels were indexed using Elasticsearch~\cite{Elasticseach2015} at a keyframe level. 
For color extraction, the dominant 8 colors, such as red, purple, dark blue light blue, green, yellow, orange, and gray, were selected to generate the 16x16 binary masks as a color feature vector~\cite{Sloth2017}. 
Similar to the color feature, we chose the top 10-object labels from 545 object classes to construct the 16x16  binary masks for spatial object extraction. 
The top 10-object labels includes of person, man, woman, face, clothing, tree, plant, car, window and poster. 
Both color and spatial feature vectors were hashed from 16x16 binary masks using locality sensitive hashes (LSHs)~\cite{Redis2013} and they were stored on the Redis.

\begin{figure}[b]
\begin{center}
\includegraphics[width=0.9\linewidth]{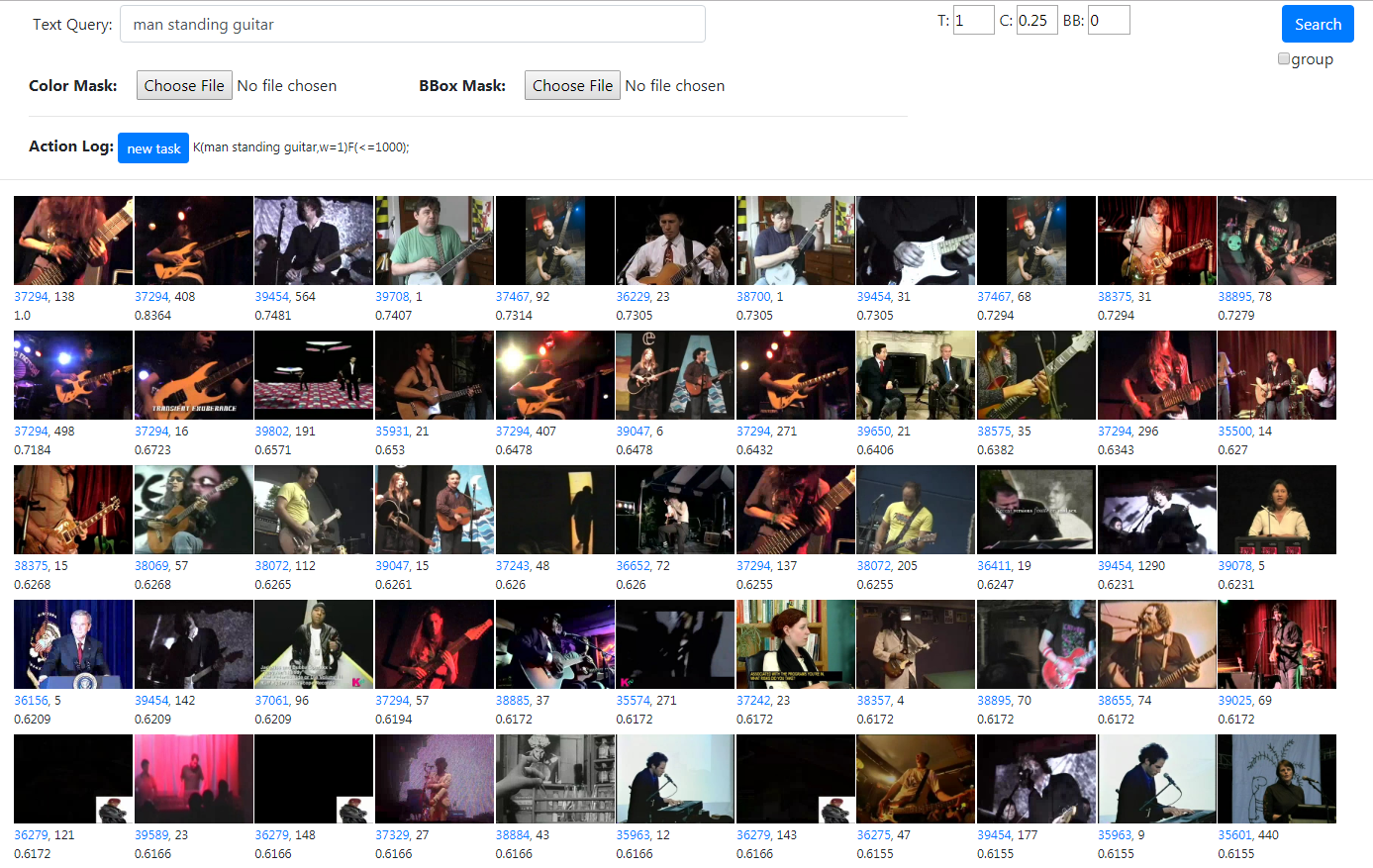}
\end{center}
\caption{Screenshot of Sloth Search System}
\label{fig:screenshot_gui}
\end{figure}

For text-based retrieval using the Elasticsearch~\cite{Elasticseach2015}, the score of an image related to the search terms were calculated based on Term Frequency/Inverse Document Frequency (TF/IDF) scoring model. 
For visual-based retrieval, the score was calculated based on the cosine distance of the hash. In order to combine the results from textual and visual search, the weighed formulation is used in Eq.~\ref{eq:weight_dist}.

\begin{equation}
\label{eq:weight_dist}
sim_{all}=w_{t} \cdot sim_{t}+w_{c} \cdot (1 - dist_{c})+ w_{o} \cdot (1-dist_{o})
\end{equation}
where $sim_{all}$ is the combined similarity score, $sim_{t}$ is the similarity score from querying by text, $dist_{c}$ and $dist_{o}$ are cosine distance from querying by color sketch and object sketch, respectively. $w_t$, $w_c$ and $w_o$ are weight parameters that can be defined by a user.

\begin{figure}[tb]
\begin{center}
\includegraphics[width=0.9\linewidth]{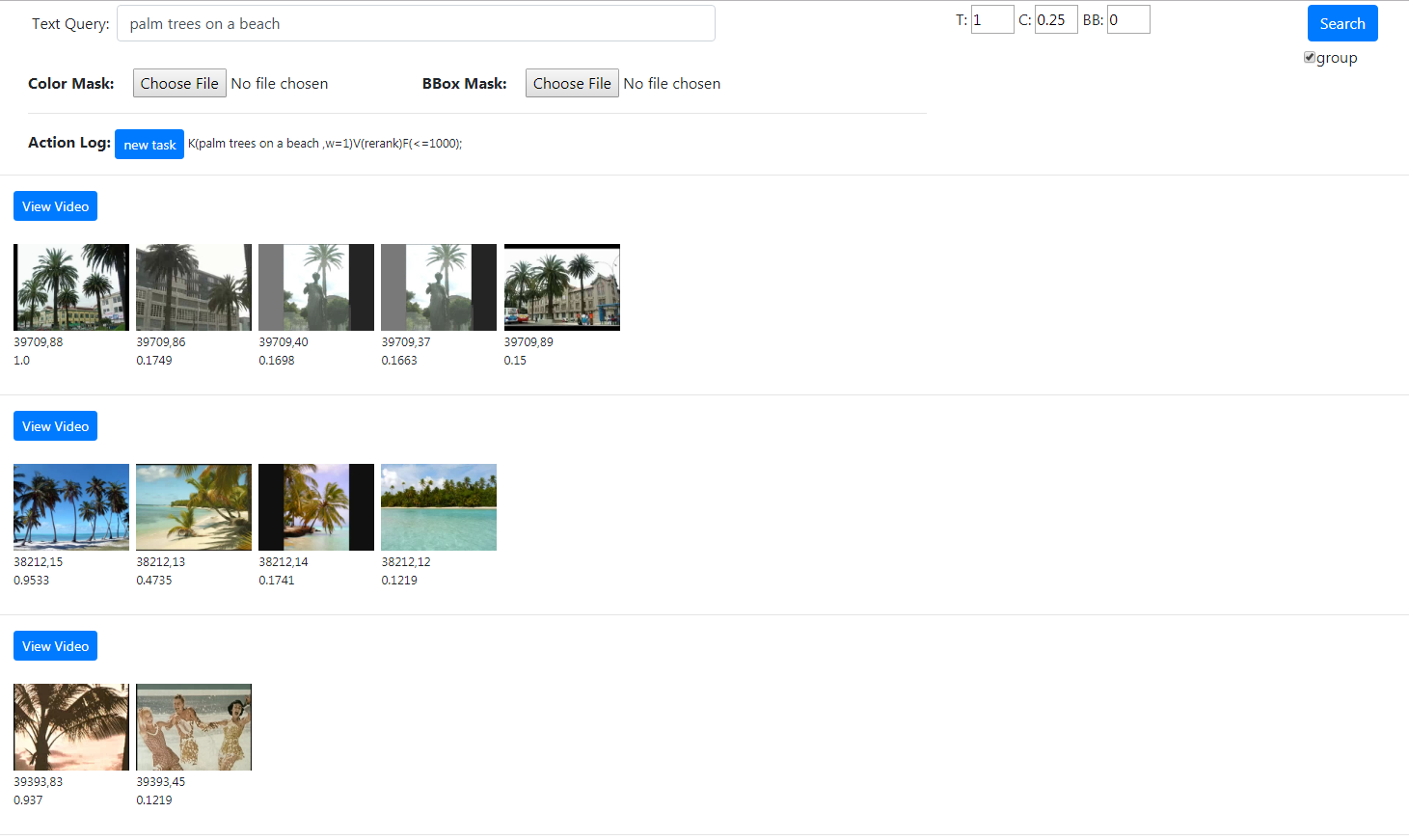}
\end{center}
\caption{Re-ranking by video grouping}
\label{fig:video_grouping}
\end{figure}

\begin{figure}[b]
\begin{center}
\includegraphics[width=0.9\linewidth]{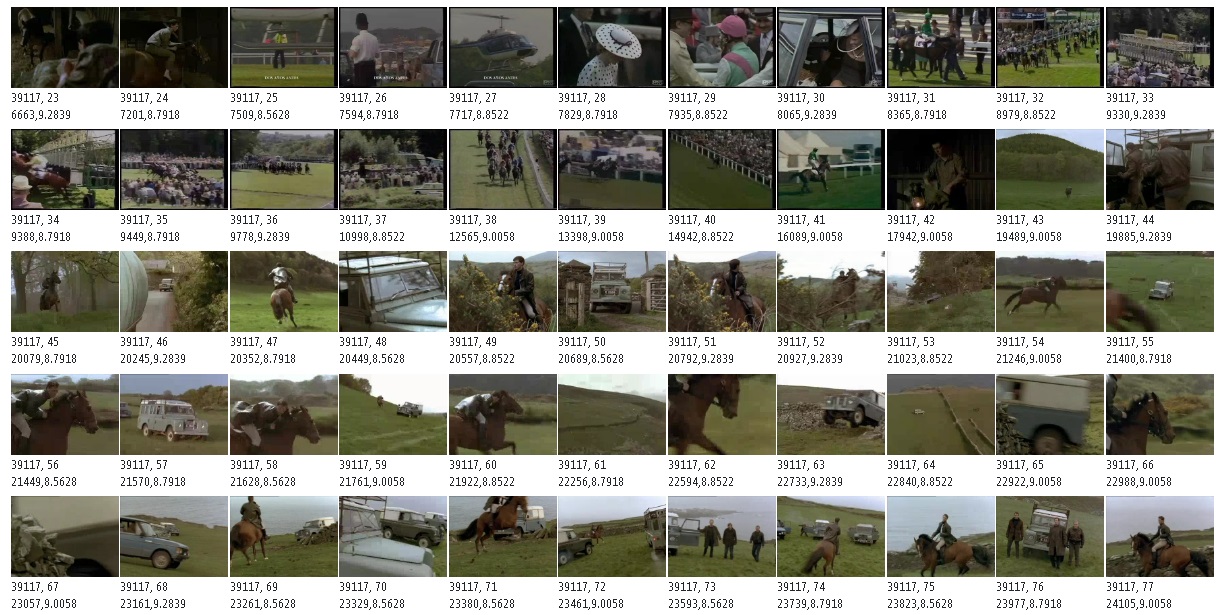}
\end{center}
\caption{Viewing all frame of each video}
\label{fig:viewall}
\end{figure}

Our tool was designed with a simple and basic interface where a user can search by entering one or multiple texts or search by uploading a sketch image, shown in Fig.~\ref{fig:screenshot_gui}. 
For visual query, the user can roughly sketch a color image or draw bounding boxes of a particular object represented by color. A combined query can be performed and the weight for each query type can be adjusted.
There are two display modes for viewing the retrieval results. For default mode, the results are displayed in grid views sorted by their similarity score, shown in Fig.~\ref{fig:screenshot_gui}. 
For grouping mode, the frames with the same video id are grouped together as a row shown in Fig.~\ref{fig:video_grouping}. And, each group is sorted according to the maximum score for each group.
Our system also have a shot view for each video, shown in Fig.~\ref{fig:viewall}. It allows a user to view all keyframes in that particular video by clicking the video id.

\section{System Usage at VBS2018}
The text-based search has been mainly used by both experts and novices during the competition. The search can be queried using multiple keywords to improve the result. This type of search is faster and easier to used compared to the visual sketch. However, the sketch-based search serves as an alternative search option when the text-based search fails. This sketch-based search has additional capabilities of color distribution and object location. 

There are three types of queries in the competition; visual KIS, textual KIS and textual AVS. For this competition, our system works well on a particular task. 
Our tool performs quite well on the AVS task. With easy-to-use browsing interface and adequate labels for general objects, the user can search for the general objects in the AVS task quite well. 
This text-based query is quite generic making our system to be able to find many relevant frames. Moreover, for a particular retrieved frame, the user has an option to view all frames in that particular video. This feature helps the user in the AVS task to retrieve more related frames from the same video.
Both expert and novice user can search and retrieve desired image for AVS task quite fast and our team was ranked first in this competition task.

For visual KIS, only two queries in the expert session were found using text-based query and none were found in the novice session. In this visual KIS task, each frame usually contains different objects making it difficult to search for that particular shot as our current system can only search for a particular frame and not a shot that composed of multiple unique objects in a different frame. Our sketch-based query also didn't perform as well as we expected and it was not able to help finding the shot when the text-based failed. 
On the other hand, textual KIS is also quite difficult as there is no visual information and we can only rely on text-based query. This textual information is usually very specific. 
Our system was not able to extract such specific detail of an object such as "a bearded man" or "a woman carries a large orange bag". Our types of object is still limited making our system unable to find an uncommon object as well.

At VBS2018 competition, our system was demonstrated to be a good interactive browsing tool. However, the system still needs some improvements. Currently, the number of keyframe is still too small and the system lacks the ability to search for multiple components in the whole shot or the attributes of an object. Reranking and filtering should be adopted for a better interactive browsing interface. And, the sketch-based search should be improved for a better result as well.



\bibliographystyle{ACM-Reference-Format}
\bibliography{sample-bibliography}


\begin{thebibliography}{7}


\ifx \showCODEN    \undefined \def \showCODEN     #1{\unskip}     \fi
\ifx \showDOI      \undefined \def \showDOI       #1{#1}\fi
\ifx \showISBNx    \undefined \def \showISBNx     #1{\unskip}     \fi
\ifx \showISBNxiii \undefined \def \showISBNxiii  #1{\unskip}     \fi
\ifx \showISSN     \undefined \def \showISSN      #1{\unskip}     \fi
\ifx \showLCCN     \undefined \def \showLCCN      #1{\unskip}     \fi
\ifx \shownote     \undefined \def \shownote      #1{#1}          \fi
\ifx \showarticletitle \undefined \def \showarticletitle #1{#1}   \fi
\ifx \showURL      \undefined \def \showURL       {\relax}        \fi
\providecommand\bibfield[2]{#2}
\providecommand\bibinfo[2]{#2}
\providecommand\natexlab[1]{#1}
\providecommand\showeprint[2][]{arXiv:#2}

\bibitem[\protect\citeauthoryear{Awad, Butt, Fiscus, and et. al}{Awad
  et~al\mbox{.}}{2017}]%
        {2017trecvidawad}
\bibfield{author}{\bibinfo{person}{George Awad}, \bibinfo{person}{Asad Butt},
  \bibinfo{person}{Jonathan Fiscus}, {and} \bibinfo{person}{et. al}.}
  \bibinfo{year}{2017}\natexlab{}.
\newblock \showarticletitle{TRECVID 2017: Evaluating Ad-hoc and Instance Video
  Search, Events Detection, Video Captioning and Hyperlinking}. In
  \bibinfo{booktitle}{\emph{Proceedings of TRECVID 2017}}. NIST, USA.
\newblock


\bibitem[\protect\citeauthoryear{Carlson}{Carlson}{2013}]%
        {Redis2013}
\bibfield{author}{\bibinfo{person}{Josiah~L. Carlson}.}
  \bibinfo{year}{2013}\natexlab{}.
\newblock \bibinfo{booktitle}{\emph{Redis in Action}}.
\newblock \bibinfo{publisher}{Manning Publications Co.},
  \bibinfo{address}{Greenwich, CT, USA}.
\newblock
\showISBNx{1617290858, 9781617290855}


\bibitem[\protect\citeauthoryear{Gormley and Tong}{Gormley and Tong}{2015}]%
        {Elasticseach2015}
\bibfield{author}{\bibinfo{person}{Clinton Gormley} {and}
  \bibinfo{person}{Zachary Tong}.} \bibinfo{year}{2015}\natexlab{}.
\newblock \bibinfo{booktitle}{\emph{Elasticsearch: The Definitive Guide}
  (\bibinfo{edition}{1st} ed.)}.
\newblock \bibinfo{publisher}{O'Reilly Media, Inc.}
\newblock
\showISBNx{1449358543, 9781449358549}


\bibitem[\protect\citeauthoryear{Krasin, Duerig, Alldrin, and et. al}{Krasin
  et~al\mbox{.}}{2017}]%
        {openimages}
\bibfield{author}{\bibinfo{person}{Ivan Krasin}, \bibinfo{person}{Tom Duerig},
  \bibinfo{person}{Neil Alldrin}, {and} \bibinfo{person}{et. al}.}
  \bibinfo{year}{2017}\natexlab{}.
\newblock \showarticletitle{OpenImages: A public dataset for large-scale
  multi-label and multi-class image classification.}
\newblock \bibinfo{journal}{\emph{Dataset available from
  https://github.com/openimages}} (\bibinfo{year}{2017}).
\newblock


\bibitem[\protect\citeauthoryear{Rujikietgumjorn, Watcharapinchai, and
  Marukatat}{Rujikietgumjorn et~al\mbox{.}}{2018}]%
        {Sloth2017}
\bibfield{author}{\bibinfo{person}{Sitapa Rujikietgumjorn},
  \bibinfo{person}{Nattachai Watcharapinchai}, {and} \bibinfo{person}{Sanparith
  Marukatat}.} \bibinfo{year}{2018}\natexlab{}.
\newblock \showarticletitle{Sloth Search System}. In
  \bibinfo{booktitle}{\emph{MultiMedia Modeling - 24th International
  Conference, {MMM} 2018, Bangkok, Thailand, February 5-7, 2018, Proceedings,
  Part {II}}}. \bibinfo{pages}{431--437}.
\newblock
\showISBNx{978-3-319-73600-6}


\bibitem[\protect\citeauthoryear{Vinyals, Toshev, Bengio, and Erhan}{Vinyals
  et~al\mbox{.}}{2017}]%
        {Vinyals2017}
\bibfield{author}{\bibinfo{person}{O. Vinyals}, \bibinfo{person}{A. Toshev},
  \bibinfo{person}{S. Bengio}, {and} \bibinfo{person}{D. Erhan}.}
  \bibinfo{year}{2017}\natexlab{}.
\newblock \showarticletitle{Show and Tell: Lessons Learned from the 2015 MSCOCO
  Image Captioning Challenge}.
\newblock \bibinfo{journal}{\emph{IEEE Transactions on Pattern Analysis and
  Machine Intelligence}} \bibinfo{volume}{39}, \bibinfo{number}{4}
  (\bibinfo{date}{April} \bibinfo{year}{2017}), \bibinfo{pages}{652--663}.
\newblock
\showISSN{0162-8828}


\bibitem[\protect\citeauthoryear{Zhou, Lapedriza, Khosla, Oliva, and
  Torralba}{Zhou et~al\mbox{.}}{2017}]%
        {zhou2017places}
\bibfield{author}{\bibinfo{person}{Bolei Zhou}, \bibinfo{person}{Agata
  Lapedriza}, \bibinfo{person}{Aditya Khosla}, \bibinfo{person}{Aude Oliva},
  {and} \bibinfo{person}{Antonio Torralba}.} \bibinfo{year}{2017}\natexlab{}.
\newblock \showarticletitle{Places: A 10 million Image Database for Scene
  Recognition}.
\newblock \bibinfo{journal}{\emph{IEEE Transactions on Pattern Analysis and
  Machine Intelligence}} (\bibinfo{year}{2017}).
\newblock


\end{thebibliography}

\end{document}